\documentclass[reprint,superscriptaddress,amsmath,amssymb,aps,pra,onecolumn]{revtex4-2}

\usepackage{subfigure}
\usepackage{tikz}
\usetikzlibrary{quantikz}
\usepackage[colorlinks,linkcolor=blue,anchorcolor=blue,citecolor=blue]{hyperref}
\usepackage{amsthm}
\usepackage{bm}
\newtheorem{mylemma}{Lemma}

\hypersetup{%
	plainpages=true,
	breaklinks=true,
	hypertexnames=false,
	pageanchor=true,
	colorlinks=true,
	linkcolor={blue},
	citecolor={blue},
	urlcolor={red},
	anchorcolor={black}
}

\newcommand{\casql}{Key Laboratory of Quantum Information, Chinese Academy of Sciences, School of Physics, University of Science and Technology of China, Hefei, Anhui, 230026, China}
\newcommand{\casex}{CAS Center For Excellence in Quantum Information and Quantum Physics, University of Science and Technology of China, Hefei, Anhui, 230026, China}
\newcommand{\hfnational}{Hefei National Laboratory, University of Science and Technology of China, Hefei, Anhui, 230088, China}
\newcommand{\aihf}{Institute of Artificial Intelligence, Hefei Comprehensive National Science Center, Hefei, Anhui, 230088, China}
\newcommand{\origin}{Origin Quantum Computing, Hefei, Anhui, 230026, China}

\begin{document}
\title{Mitigating Barren Plateaus with Transfer-learning-inspired Parameter Initializations}


\author{Huan-Yu Liu}
\email{liuhuany@mail.ustc.edu.cn}
\affiliation{\casql}
\affiliation{\casex}
\affiliation{\hfnational}

\author{Tai-Ping Sun}
\affiliation{\casql}
\affiliation{\casex}
\affiliation{\hfnational}

\author{Yu-Chun Wu}
\email{wuyuchun@ustc.edu.cn}
\affiliation{\casql}
\affiliation{\casex}
\affiliation{\hfnational}
\affiliation{\aihf}

\author{Yong-Jian Han}
\affiliation{\casql}
\affiliation{\casex}
\affiliation{\hfnational}
\affiliation{\aihf}

\author{Guo-Ping Guo}
\affiliation{\casql}
\affiliation{\casex}
\affiliation{\hfnational}
\affiliation{\aihf}
\affiliation{\origin}

\date{\today}

\begin{abstract}
  Variational quantum algorithms (VQAs) are widely applied in the noisy intermediate-scale quantum era and are expected to demonstrate quantum advantage.
  However, training VQAs faces difficulties, one of which is the so-called barren plateaus (BP) phenomenon, where gradients of cost functions vanish exponentially with the number of qubits.
  In this paper, inspired by transfer learning, where knowledge of pre-solved tasks could be further used in a different but related work with training efficiency improved, we report a parameter initialization method to mitigate BP.
  In the method, a small-sized task is solved with a VQA. Then the ansatz and its optimum parameters are transferred to tasks with larger sizes.
  Numerical simulations show that this method could mitigate BP and improve training efficiency.
  A brief discussion on how this method can work well is also provided.
  This work provides a reference for mitigating BP, and therefore, VQAs could be applied to more practical problems.
\end{abstract}

\maketitle

\section{Introduction}
While the realization of quantum computers that can carry out fault-tolerant quantum computations \cite{FTQC1} will potentially take decades of research, designing algorithms executable with noisy intermediate-scale quantum (NISQ) devices \cite{nisq1,nisq2} to solve practical problems is imminent. Variational quantum algorithms (VQAs) \cite{vqa1,vqa2,vqa3} are promising in this aspect, which have been applied in various regions, including simulation of molecular systems in quantum chemistry \cite{vqa3,vqe1,vqe2,vqe3}, scientific computation \cite{vqls1,vqde,vqpe}, machine learning \cite{qml,qml2,qml3}, system-environment interaction simulation \cite{vqoqs1,vqoqs2}, etc.
Compared to traditional quantum algorithms like quantum phase estimation \cite{pea}, VQAs have a weaker requirement for the qubit number and quantum operation depth. But it contains a cost of training parameterized quantum circuits (PQCs).

Training VQAs has been proven to be NP-hard \cite{vqahard}. And recent works reported that such a training process can suffer from the barren plateaus (BP) phenomenon \cite{bp,bplocal}, where gradients of cost functions vanish exponentially with the number of qubits. Therefore, exponential shots would be needed to sample one gradient, which limits applications of VQAs to larger-scaled problems.

\begin{figure*}[ht]
  \centering
  \includegraphics[width=0.5\linewidth]{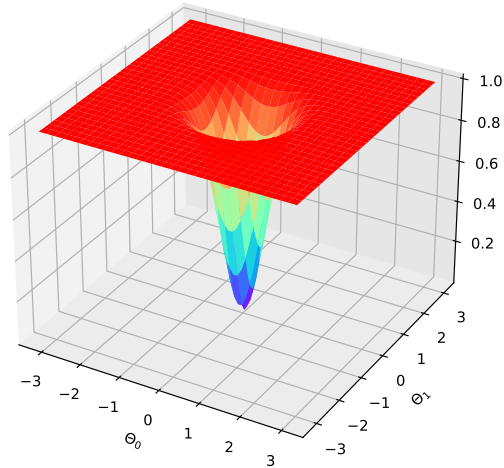}
  \caption{
  \textbf{A cost landscape when BP exists}.
  Parameters in the BP region have vanishing gradients.
  Only parameters in the exponentially suppressed central part can be optimized efficiently.
  }\label{bpfig}
\end{figure*}

Fig. \ref{bpfig} shows a cost landscape with BP. Parameters in the BP region (red part) have vanishing gradients. Only parameters in the central part, which is exponentially suppressed, can be efficiently optimized. Works to deal with BP have been proposed: A shallow PQC will not suffer from BP with a local Hamiltonian \cite{bplocal}. BP is absent with quantum convolution neural networks and tree tensor networks \cite{bpzx}. Determining a good set of initial parameters also works \cite{bpini,llbp}.

Transfer learning \cite{tl1,tl3,tlnew} is a widely applied parameter initialization strategy in machine learning. It tries to apply information of pre-solved tasks (base task) to a different but related task (target task). It is more efficient with applications of that knowledge in some cases \cite{tleff,tl4}.

In this paper, inspired by transfer learning, we propose a parameter initialization method to mitigate BP. A small-sized task is trained as the base task. Then the PQC and its optimum parameters are transferred to target tasks with larger sizes. Numerical simulations show that this method can mitigate BP and improve training efficiency. A brief discussion on the feasibility of this method in general cases is also provided, showing that this method could work well, especially when solving translationally invariant systems.

The rest of this paper is organized as follows:
Sec. \ref{background} mainly introduces variational quantum algorithms, ansatzes and BP.
Sec. \ref{method} gives methods for obtaining initial parameters for PQCs and the relation between some existing works.
Simulation results to show the performance of our method are shown in Sec. \ref{simulation}.
In Sec. \ref{discuss}, we provide discussions on how our method could work well and a conclusion is given in Sec. \ref{conclusion}.

\section{preliminaries}\label{background}

\subsection{Variational quantum algorithms}
A VQA is a hybrid quantum-classical algorithm that employs a quantum processor for ansatz preparations and expectation value measurements and a classical optimizer for parameter optimizations. Generally, an ansatz is generated as $|\psi(\bm{\theta})\rangle=U(\bm{\theta})|0\rangle$ with the PQC $U(\bm{\theta})$ having the form:
\begin{equation}\label{PQC}
  U(\bm{\theta}) = \prod_{l=1}^{L} U_l(\theta_l),
\end{equation}
where $U_l(\theta_l)=\exp ( -i\theta_l V_l /2 )W_l$ with $V_l^2=I$, which is usually a tensor product of Pauli operators. $W_l$ is an un-parameterized quantum gate and $\bm{\theta}\equiv \{ \theta_l \}_{l=1}^L$ is the set of parameters to be optimized. The cost function in a VQA is usually an expectation value of Hamiltonian $H$ corresponding to the problem:
\begin{equation}\label{cost}
C(\bm{\theta}) = \operatorname{Tr}[     HU(\bm{\theta})\rho_0 U(\bm{\theta})^{\dagger}    ],
\end{equation}
with $\rho_0$ the initial state. Generalized cost functions can be functions of measured expectation values, which can be efficiently evaluated with a classical computer.

Ansatzes in VQAs can be either problem-agnostic or problem-inspired. Problem-agnostic ansatzes do not depend on any information about the problem to be solved. Strong expressibility \cite{expmpqc,exp1,exp2,exp3} is always needed and a widely used example is hardware-efficient ansatz \cite{vqe1,hea2}. In contrast, problem-inspired ansatzes try to include known information about the problem to be solved, which confine their search space to smaller scopes. Examples include unitary coupled-cluster ansatz \cite{ucc1,ucc2} in quantum chemistry and Hamiltonian variational ansatz \cite{hva1,hva2}.

Parameter optimizations can be accomplished by gradient-based or gradient-free methods. Gradient-based algorithms optimize parameters based on sampled gradients. A commonly applied one is the BFGS algorithm \cite{bfgs}. Gradient-free method updates parameters based on sampled cost differences. Ref. \cite{optimize} tested the performance of solving VQAs with a wide range of optimization methods.

\begin{figure*}[ht]
\centering
\subfigure[HEA]{
\begin{tikzpicture}
  \node[scale=0.65]{
  \begin{tikzcd}
    \qw & \gate[wires=4]{\text{HEA}} & \qw \\
    \qw &                     & \qw \\
    \qw &                     & \qw \\
    \qw &                     & \qw
  \end{tikzcd}=
  \begin{tikzcd}
    \qw & \gate[wires=4,style={fill=red!20},label style=cyan]{U_{\text{single}}(\bm{\theta}_p)}\gategroup[4,steps=2,style={dashed,
rounded corners,fill=blue!20, inner xsep=2pt},
background]{{ repeat this block}} & \gate[wires=4,style={fill=green!20},label style=cyan]{U_{\text{ent}}} & \qw \\
    \qw &         &               & \qw \\
    \qw &          &              & \qw \\
    \qw &         &               & \qw
  \end{tikzcd}
  };
\end{tikzpicture}
}
\subfigure[HVA]{
\begin{tikzpicture}
  \node[scale=0.65]{
  \begin{tikzcd}
    \qw & \gate[wires=4]{\text{HVA}} & \qw \\
    \qw &                     & \qw \\
    \qw &                     & \qw \\
    \qw &                     & \qw
  \end{tikzcd}=
  \begin{tikzcd}
    \qw & \gate[wires=4,style={fill=red!20},label style=cyan]{ e^{ -i\theta_{p,1} H_1  } }\gategroup[4,steps=3,style={dashed,
rounded corners,fill=blue!20, inner xsep=2pt},
background]{{ repeat this block}} & \gate[wires=4,style={fill=green!20},label style=cyan]{ e^{ -i\theta_{p,2} H_2  } } & \gate[wires=4,style={fill=yellow!20},label style=cyan]{ e^{ -i\theta_{p,3} H_3  } } & \qw \\
    \qw &         &        &             & \qw \\
    \qw &          &        &            & \qw \\
    \qw &         &         &            & \qw
  \end{tikzcd}
  };
\end{tikzpicture}
}
\subfigure[Network transfer]{
\begin{tikzpicture}
  \node[scale=0.65]{
  \begin{tikzcd}
    \qw  & \gate[wires=2]{U(\theta^*)}  & \qw \\
    \qw  &                    & \qw
  \end{tikzcd}$\rightarrow$
  \begin{tikzcd}
    \qw & \gate[wires=2,style={fill=yellow!20},label style=cyan]{U(\theta^*)}  &\qw &\qw &\qw \\
    \qw &  & \gate[wires=2,style={fill=red!20},label style=cyan]{U(\theta_{r})} &\qw &\qw \\
    \qw &\qw &    & \gate[wires=2,style={fill=yellow!20},label style=cyan]{U(\theta^*)} &\qw \\
    \qw &\qw &\qw &  &\qw
  \end{tikzcd}
  };
\end{tikzpicture}
}
\subfigure[Structure transfer of a HEA]{
\begin{tikzpicture}
  \node[scale=0.7]{
  \begin{tikzcd}
    \qw & \gate[wires=2]{U_{\text{single}}(\theta)} & \gate[wires=2]{U_{\text{ent}}} & \gate[wires=2]{U_{\text{single}}(\theta)} & \gate[wires=2]{U_{\text{ent}}} &\qw \\
    \qw & & & & & \qw
  \end{tikzcd}$\rightarrow$
  \begin{tikzcd}
    \qw &\gate[wires=2]{U_{\text{single}}(\theta)}\gategroup[2,steps=4,style={dashed,
rounded corners,fill=blue!20, inner xsep=2pt},
background]{{ transfer blocks}} & \gate[wires=4]{U_{\text{ent}}} &\gate[wires=2]{U_{\text{single}}(\theta)} & \gate[wires=4]{U_{\text{ent}}} &\gate[wires=2]{U_{\text{single}}(\theta)} & \gate[wires=4]{U_{\text{ent}}} &\gate[wires=2]{U_{\text{single}}(\theta)} & \gate[wires=4]{U_{\text{ent}}} &\qw \\
    \qw &                          & &                          & &                          & &                          & &\qw \\
    \qw &\gate[wires=2]{U_{\text{single}}(\theta)} & &\gate[wires=2]{U_{\text{single}}(\theta)} & &\gate[wires=2]{U_{\text{single}}(\theta)} & &\gate[wires=2]{U_{\text{single}}(\theta)} & &\qw \\
    \qw &                          & &                          & &                          & &                          & &\qw
  \end{tikzcd}
  };
\end{tikzpicture}
}
\caption{
\textbf{Ansatzes and transfer methods in this work.}
(a) A 4-qubit HEA. Each block is comprised of single- and two-qubit operations, which are defined in Eq. (\ref{headec}). The block is repeated to improve the expressibility.
(b) A 4-qubit HVA with a 3-part Hamiltonian. Each layer consists of some $e^{ -i\theta_{p,m} H_m  }$ defined in Eq. (\ref{hav}).
(c) Network transfer from 2 qubits to 4 qubits.  $\theta^*$ indicates that initial parameters are from the trained base circuit and $\theta_r$ means randomly initialized parameters.
(d) Structure transfer from a 2-qubit 2-layer HEA to a 4-qubit 4-layer one. Each transfer block, which is specified by a dashed line, can be either copied from the base circuit or randomly initialized.
}\label{123}
\end{figure*}
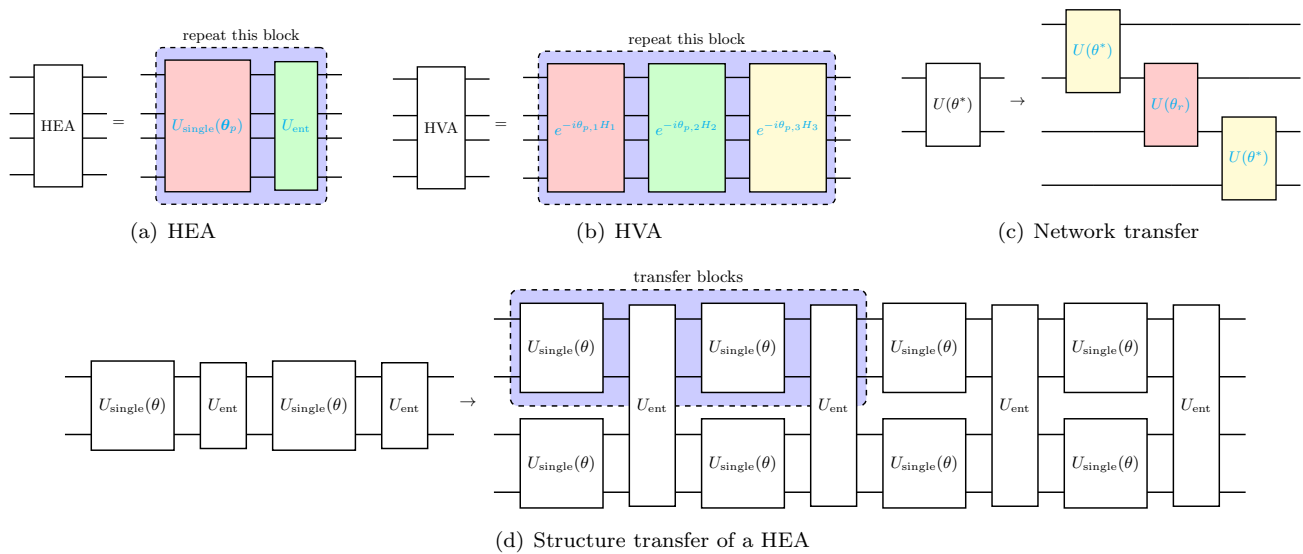

Below we will introduce ansatzes used in this work:

\paragraph{Hardware-efficient ansatz (HEA)}
Generally, a HEA is comprised of many blocks, each of which consists of single- and two-qubit gates that are easily implementable with current quantum processors. The circuit of the HEA used in this work is:
\begin{equation}\label{hea}
  U(\bm{\theta}) = \prod_{p=1}^P U_{\text{ent}} U_{\text{single}}(\bm{\theta}_p),
\end{equation}
where $P$ is the number of blocks. $U_{\text{single}}(\bm{\theta}_p)$\text{ and} $U_{ent}$ are single-qubit rotation and two-qubit entangling operations, respectively. Detailed quantum gate sequences of these operations are:
\begin{equation}\label{headec}
  \begin{aligned}
   U_{\text{single}}(\bm{\theta}_p) &=\prod_{i=1}^{n} R^i_z(\theta_{pi1}) R^i_x(\theta_{pi2}) R^i_z(\theta_{pi3}),   \\
   U_{\text{ent}} &= \left[  \prod_{i=1}^{n-1} CZ_{i,i+1}  \right] CZ_{n,1}.
  \end{aligned}
\end{equation}
Here, $n$ refers to the number of qubits. $R_{\alpha}^i(\beta)$ is a rotation gate with an angle $\beta$ in the $\alpha$-axis acting on the $i$-th qubit and $CZ= \operatorname{diag}\{1,1,1,-1\}$ is the Controlled-Z gate. The expressibility of HEAs has been shown to overpass Boltzmann machine and tensor networks \cite{expmpqc}. The quantum circuit for a 4-qubit HEA is shown in Fig. \ref{123}(a).

\paragraph{Hamiltonian variational ansatz (HVA)} The HVA originates from adiabatic quantum computing, which uses operators in the Hamiltonian to generate an ansatz. Specifically, for the Hamiltonian represented as a linear combination of operators: $H=\sum_{m=1}^M H_m$, the HVA has the form:
\begin{equation}\label{hav}
  U(\bm{\theta})  = \prod_{p=1}^P \left(  \prod_{m=1}^M  e^{ -i\theta_{p,m} H_m  }    \right),
\end{equation}
where $P$ and $M$ are the number of layers of the HVA and the number of terms of the Hamiltonian, respectively. The initial state should have a non-zero overlap with the ground state of $H$ (we can prepare it as the ground state of some $H_m$). Similar to the quantum alternating operator ansatz (QAOA), the performance depends on its number of layers \cite{qaoalayer}. The performance is also related to the interaction distance to target states \cite{QAOADIS}. The quantum circuit for a 4-qubit HVA is shown in Fig. \ref{123}(b).

\subsection{Barren plateaus and cost concentration}
Gradient-based optimization algorithms update parameters based on gradient information $\nabla C\equiv\nabla C(\bm{\theta})$, which is a vector of partial derivatives of the cost function:
\begin{equation}\label{gradient}
  \nabla C = \{ \partial_lC  \}_{l=1}^L,  \qquad \partial_lC = \frac{\partial C}{\partial\theta_l}.
\end{equation}

BP refers to the situation that $\nabla C$ vanishes exponentially with the number of qubits:
\begin{mylemma}[\textbf{Barren plateaus}]\label{bpl}
  Consider the ansatz and cost function introduced in Eq. (\ref{PQC}) and Eq. (\ref{cost}). Denote: $U = U_LU_jU_R$, where $U_L =\prod_{l<j} U_l$ and $U_R = \prod_{l>j} U_l$. Then we have:
  \begin{itemize}
    \item The average of the partial derivative at any parameter over the ansatz is 0: $\langle \partial_jC\rangle=0, \forall j$.
    \item The variance of the above partial derivative vanishes exponentially with the number of qubits $n$: $\operatorname{ Var}[\partial_jC]\in O(p^{-n}),p> 1$, when either $U_L$ or $U_R$ forms a unitary 2-design.
  \end{itemize}
\end{mylemma}

Proofs of this lemma can be found in \cite{bp,expbp}. A circuit $U$ forms a unitary 2-design \cite{u2d1,u2d2} indicates that average of the function $f(U)$ up to the second order of $U$ and $U^{\dagger}$ (An example of such a function is $\operatorname{Tr}[UAU^{\dagger}B] \operatorname{Tr}  [UCU^{\dagger}D]$ with $A,B,C,D$ Hermite operators, which will be used when analysing  $\operatorname{Var}[\partial_jC]$) over such circuit is indistinguishable from average of that function over the unitary group via the unitarily invariant Haar measure. Since $\langle \partial_jC\rangle = 0$, then $\operatorname{Var}[\partial_jC] = \langle (\partial_jC)^2 \rangle = \int dU (\partial_jC)^2 $. The integral can be replaced by the Weingarten functions \cite{wf} if $U$ forms a unitary 2-design. The result, therefore, can be evaluated efficiently, which establishes an exponential decay with the number of qubits.

According to Chebyshev's inequality:
\begin{equation}\label{Chebyshev}
  P(  |\partial_jC |\geq c ) \leq \frac{ \operatorname{ Var}[ \partial_jC ]  }{c^2},
\end{equation}
when the variance of $\partial_jC$ establishes an exponential decay, the probability of finding a $|\partial_jC|$ that is bigger than $c$ also decreases exponentially. This indicates that one needs exponential shots to sample the gradient information. Therefore, optimizing parameters in the BP regions with gradient-based methods is difficult.

The expressibility of a quantum circuit is defined based on the ``distance" from the circuit to form a unitary 2-design \cite{exp1,exp2}. It was shown in \cite{expbp} that an ansatz with strong expressibility, i.e, close to forming a unitary 2-design (We also call it forms an approximate unitary 2-design), would have poor trainability. While it is shown that random circuits are approximate 2-design \cite{rqc2d}, it indicates that HEA, which is a non-structured circuit, will establish BP. It was also shown in \cite{expbp} that restricting the expressibility of an ansatz, like correlating parameters, can reduce the speed of decay of gradient.

From the above analysis with Chebyshev's inequality, BP also establishes a cost concentration \cite{bpcostconcentration}, where cost functions are exponentially concentrated around their mean value. This phenomenon can be represented using the variance of cost difference similar to Lemma \ref{bpl}:
\begin{equation}\label{costconcentration}
\operatorname{Var} [  C(\bm{\theta}) -\langle C \rangle   ] \in O(b^{-n}), b>1.
\end{equation}

This means that when BP exits, the probability of finding a set of parameters whose cost function is lower than the average value at a constant c is also exponentially suppressed. This exponential suppression of cost differences makes parameter optimization also difficult with gradient-free methods \cite{bpgrafree}, where parameters are updated based on sampled cost differences. It has been shown that  BP is equivalent to cost concentration \cite{bpcostconcentration}.

\section{Method}\label{method}

According to the former analysis, when BP exists, training with randomly initialized parameters will be difficult. From this point, a good set of initial parameters could be significant. If initial parameters are located out of the BP region, training could get started. Training efficiency could also be improved compared to randomly initial parameters.

Inspired by transfer learning, we propose a parameter initialization method to mitigate BP. A small-sized task is trained as the base task. Then the structure and optimum parameters of the base circuit are transferred to target tasks with larger sizes. Simulations with that transferred set of initial parameters and randomly initialized parameters are called hot start and cold start, respectively. Below we will introduce two methods to transfer the information.

\begin{enumerate}
  \item \textbf{Network transfer: } The network transfer method generates a target circuit by simply copying the base circuit. Specifically, suppose an $n$-qubit base circuit $U(\bm{\theta})_{1,n}$ is trained with optimum parameters $\bm{\theta}^*$, where subscripts $``1,n"$ means that the circuit acts on ranging from the first to the $n$-th qubit. We can then obtain the target circuit $\mathcal{U}(\bm{\theta})$ by arranging the base circuit in  an alternating layer if we were to solve the target task with $m(m>n)$ qubits:
\begin{equation}\label{nettransfer}
  \mathcal{U}(\bm{\theta}) = \prod_{i=1}^{m-n+1} U(\bm{\theta}_i)_{i,i+n-1},
\end{equation}
where each $\bm{\theta}_i$ can be either $\bm{\theta}^*$ or randomly initialized $\bm{\theta}_r$, which is represented by T(transfer) and R(random). Finally, this parameter initialization will be identified by a string like ``TRT$\cdots$". Note that this method is only suitable for problem-agnostic ansatzes. This alternating arrangement of problem-inspired ansatzes will not preserve any physical information. A schematic representation for the network transfer method is shown in Fig. \ref{123}(c).
  \item \textbf{Structure transfer: }Different from the network transfer method, the structure transfer method keeps the structure of the circuit unchanged in the transfer process. All parameters in the target circuit are divided into several parts to be transferred, which will be identified by a string comprised of T and R similar to the network transfer method. An example of the structure transfer method for HEA is shown in Fig. \ref{123}(d). As it preserves the structure of the ansatz, both problem-inspired and problem-agnostic ansatzes can be transferred with this method.
\end{enumerate}

Previous works have been focusing on determining a good set of initial parameters. We summarized and discussed two of them here.

\begin{enumerate}
  \item \textbf{Block identity encoding (BLE): } Ref. \cite{bpini} provides an initialization strategy for PQCs. For a PQC to be trained, they randomly choose some parameters and adjust the others such that the PQC becomes a sequence of shallow blocks, each of which compiles to an Identity. This will reduce the effective depth and break the unitary 2-design structure, which enables it to be trained.

      Note that this method has a special requirement on the structure of an ansatz and is hard to be implemented in general. The main reason is that for one layer of the ansatz with some parameters fixed, it is difficult to compute the rest such that the block compiles to an identity. Moreover, the strategy can be applied in our method as well. In the transfer process, there exist some randomly initialized parameters,  identified by ``R". We can adjust these parameters according to the BLE method.

  \item \textbf{Layerwise learning: } Ref. \cite{llbp} proposes a training method as well as a parameter-initialization process. They first grow the ansatz by layers, where mainly newly-added parameters are trained. After the final ansatz is obtained, this pre-training will provide a specific set of parameters instead of randomly initialized ones. Then the whole network is trained with their method.

      Note that this work can be viewed as a special case when the size of the base network is equal to the target one. And the training strategy introduced in that work can also be applied in our method.
\end{enumerate}

\section{numerical simulations}\label{simulation}
Numerical simulations are performed in this part as benchmarks. We consider solving the ground states of some systems, which is a main topic of VQAs. We apply the BFGS algorithm for parameter optimizations, which is commonly considered. Ansatzes and transfer methods introduced before will be applied in following simulations. Both base and target circuits will be trained to obtain 100 successful runs, where a successful run means that the optimum ground state energy is within chemical accuracy ($1.6\times 10^{-3}$ Hartree) compared to the exact value. In all models, the simulation starts with training a base circuit with randomly initialized parameters. meanwhile recording the set of optimum parameters in each successful run. These optimum parameters will form a pool and will be randomly picked out for simulations of target tasks. Target circuits with different parameter initialization methods will be simulated. We will introduce tasks considered first, then simulation results will be given to show the performance of different parameter initialization methods.
\subsection{Tasks}
\begin{itemize}
  \item The transverse field Ising model, whose Hamiltonian is:
  \begin{equation}
  H = -J_{\text{Ising}} \sum_{ \{i,j\} } Z_iZ_j - h \sum_i X_i,
  \end{equation}
  where $P_j(P=X/Y/Z)$ is the Pauli operator on the $j$-th qubit. $\{i,j\}$ represents the nearest iteration pairs. $J_{\text{Ising}}$ and $h$ model the spin-coupling strength in the $z$-axis and the transverse field along the $x$-axis, respectively. We perform numerical simulations on the one-dimensional model with $J_{\text{Ising}}=1$ and $h=2$. We define \textbf{Task A} and \textbf{Task B} to solve the 6-qubit and 8-qubit models, respectively. A 4-qubit model is trained as the base task for these two tasks. A 4-layer HEA and the network transfer method are applied.
  \item Hydrogen atoms in a linear structure with atom-atom distance at 0.74\AA. We consider STO-3G basis and the Jordan-Wigner transformation \cite{jw}, which results in a spin Hamiltonian of $2m$ qubits for $m$ atoms, expressed as $H=\sum_Ic_IP_I$, with $c_I$ a real coefficient and $P_I$ a tensor product of Pauli operators. We consider \textbf{Task C}: The base task is to solve $H_2$ with a 4-qubit 4-layer HEA. The target task is to solve $H_3$ with a 6-qubit 8-layer HEA. The structure transfer method is applied and the top 4 qubits are transferred.
  \item The Heisenberg XXZ model:
  \begin{equation}\label{hamxxz}
  H = -J_{\text{XXZ}} \sum_{ \{i,j \}  } X_iX_j +Y_iY_j +\Delta Z_iZ_j,
  \end{equation}
  where $J_{\text{XXZ}}$ represents the in-plane spin coupling strength and $\Delta$ models the dimensionless anisotropy factor between the $xy$ plane and $z$ axis coupling strength. In numerical simulations, we set $J_{\text{XXZ}}=1$ and $\Delta=2$. The base task is a 4-qubit one-dimensional model. For target tasks, we consider both the one- and two-dimensional models, which are denoted as \textbf{Task D} and \textbf{Task E} respectively.

  For the one-dimensional target model, we consider the HVA and the structure transfer method. The Hamiltonian can be divided into three parts as $H=H_X+H_Y+H_Z$, which results in three parameters per layer. The number of layers for the base and the target circuit is 4 and 8, respectively.  For the two-dimensional model with size $2\times 4$, we apply the HEA and the structure transfer method. The number of layers of ansatzes is the same as the one-dimensional case.
  \item \textbf{Task F} evaluates the performance between our method and the BLE strategy introduced before. We consider the one-dimensional XXZ model defined in Eq. (\ref{hamxxz}). Model parameters are the same as before. To better implement the BLE strategy, we apply a variant of the HVA, where every two layers can compile to an identity:
  \begin{equation}\label{hvavar}
  U(\bm{\theta}) = \prod_{p=1}^{P} V_p(\bm{\theta}_p), \qquad V_p(\bm{\theta}_p) = \begin{cases}
     \prod_{m=1}^M  e^{ -i\theta_{p,m} H_m  } , & p=1,3,5,\cdots,  \\
     \prod_{m=M}^1  e^{ -i\theta_{p,m} H_m  } , & p=2,4,6,\cdots.
  \end{cases}
  \end{equation}
  In our method, we first train a 4-qubit model with a 4-layer ansatz as the base task. An 8-qubit model is chosen as the target task and we use an 8-layer ansatz as the target circuit. The former and latter 4 layers can be either randomly initialized or transferred from the base circuit. In the BLE strategy, we randomly initialize $\bm{\theta}_p=\{\theta_{p,m}\}_{m=1}^M$ when $p$ is odd. Then parameters in the $p+1$ layer are set as: $\bm{\theta}_{p+1} =\{\theta_{p+1,m}| \theta_{p+1,m} = -\theta_{p,m} \}_{m=M}^1 $ such that $V_p(\bm{\theta}_p)V_{p+1}(\bm{\theta}_{p+1})=I$.
\end{itemize}

\subsection{Simulation results}
Besides the concept of ``successful run" introduced before, training a set of parameters may fail in finding the ground state with desired accuracy. Therefore, we first consider the total trial number (TTN) to achieve 100 successful runs, which quantifies the success probability. A smaller TTN indicates a higher success probability, which is preferred. Such an indicator can be affected by BP, local minimum points, etc.

TABLE \ref{TTN} lists total trial numbers for different parameter initialization methods in every task. Transfer-learning-inspired parameters improved the performance compared to cold start parameters in all tasks. The BLE strategy does not provide better performance compared to randomly initialized ones in \textbf{Task F}. Results showed that our method can improve training efficiency.

\begin{table}[ht]
    \centering
	\caption{
    \textbf{Total trial numbers for different tasks.} In every sub-table, the first column refers to parameter initialization methods. The second column represents TTNs. Transfer-learning-inspired initial parameters increased the success probability in every model compared to cold start ones and the BLE strategy.
    }\label{TTN}
    \subfigure[\textbf{Task A}]{
    \begin{tabular}{c|c}
      \hline
      method & TTN \\
      \hline
      TTT & 101 \\
      RRT & 105 \\
      TTR & 113 \\
      RRR & 115 \\
      \hline
    \end{tabular}
    }
    \subfigure[\textbf{Task B}]{
    \begin{tabular}{c|c}
      \hline
      method & TTN \\
      \hline
      TTTTT & 105 \\
      TRTRT & 103 \\
      RTRTR & 102 \\
      RRRRR & 111 \\
      \hline
    \end{tabular}
    }
    \subfigure[\textbf{Task C}]{
    \begin{tabular}{c|c}
      \hline
      method & TTN \\
      \hline
      TT & 372 \\
      TR & 326 \\
      RT & 291 \\
      RR & 634 \\
      \hline
    \end{tabular}
    }
    \subfigure[\textbf{Task D}]{
    \begin{tabular}{c|c}
      \hline
      method & TTN \\
      \hline
      TT & 160 \\
      TR & 130 \\
      RT & 138 \\
      RR & 240 \\
      \hline
    \end{tabular}
    }
    \subfigure[\textbf{Task E}]{
    \begin{tabular}{c|c}
      \hline
      method & TTN \\
      \hline
      TTTT & 100 \\
      TRRT & 100 \\
      RTTR & 100 \\
      RRRR & 102 \\
      \hline
    \end{tabular}
    }
    \subfigure[\textbf{Task F}]{
    \begin{tabular}{c|c}
      \hline
      method & TTN \\
      \hline
      TT & 112 \\
      TR & 126 \\
      RT & 128 \\
      RR & 143 \\
      BLE & 144 \\
      \hline
    \end{tabular}
    }
\end{table}

Besides the total trial number, the number of iterations in the parameter optimization process is another indicator for identifying training efficiency. Therefore, we consider the average iteration times in the parameter optimization process with the gradient-based method over the 100 successful runs. This is a direct comparison between the hot and cold start simulations, where a smaller number indicates that the set of initial parameters is better than those with more iteration numbers.

\begin{figure}[ht]
  \centering
  \subfigure[\textbf{Task A}]{
  \includegraphics[width=0.3\linewidth]{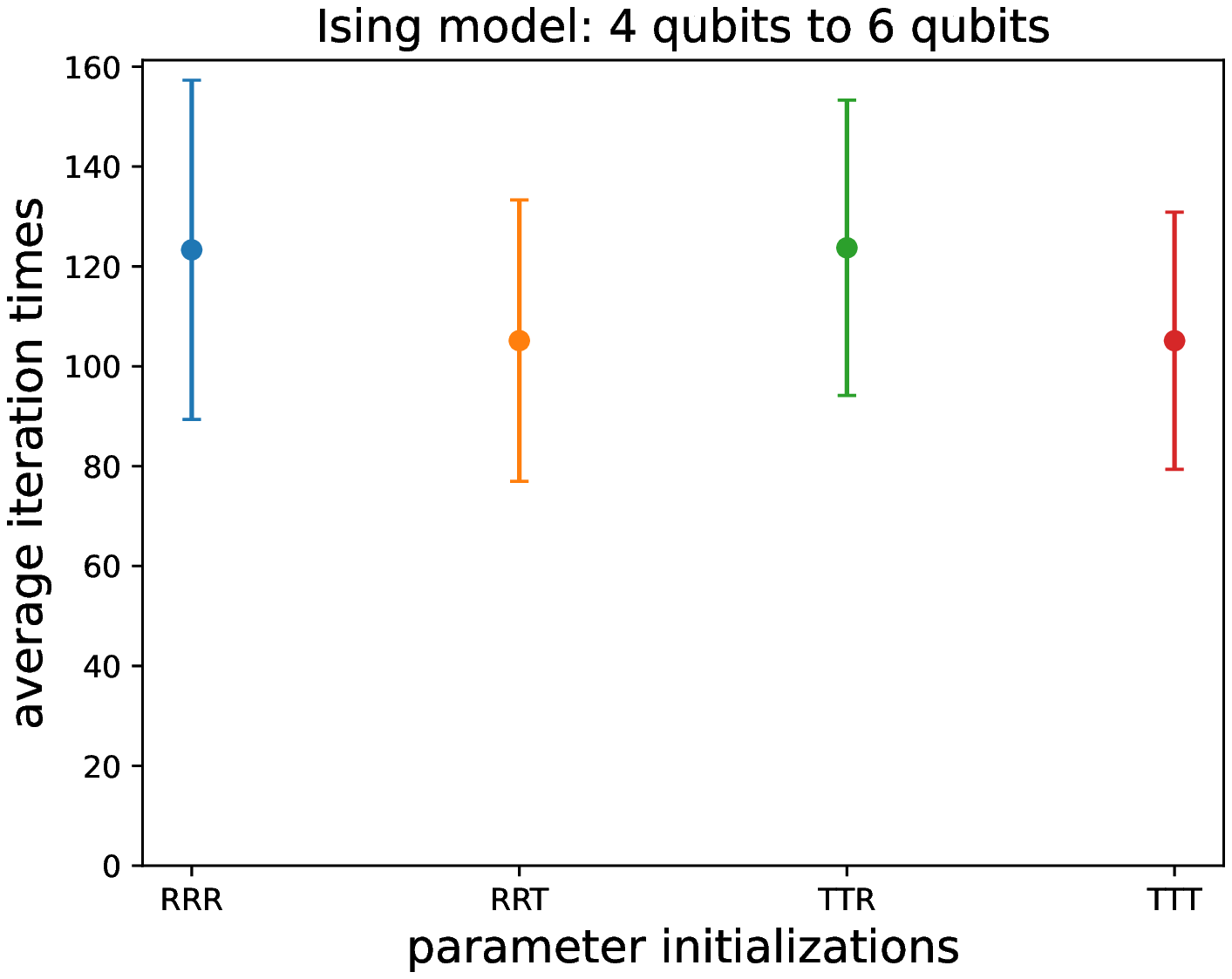}
  }
  \subfigure[\textbf{Task B}]{
  \includegraphics[width=0.3\linewidth]{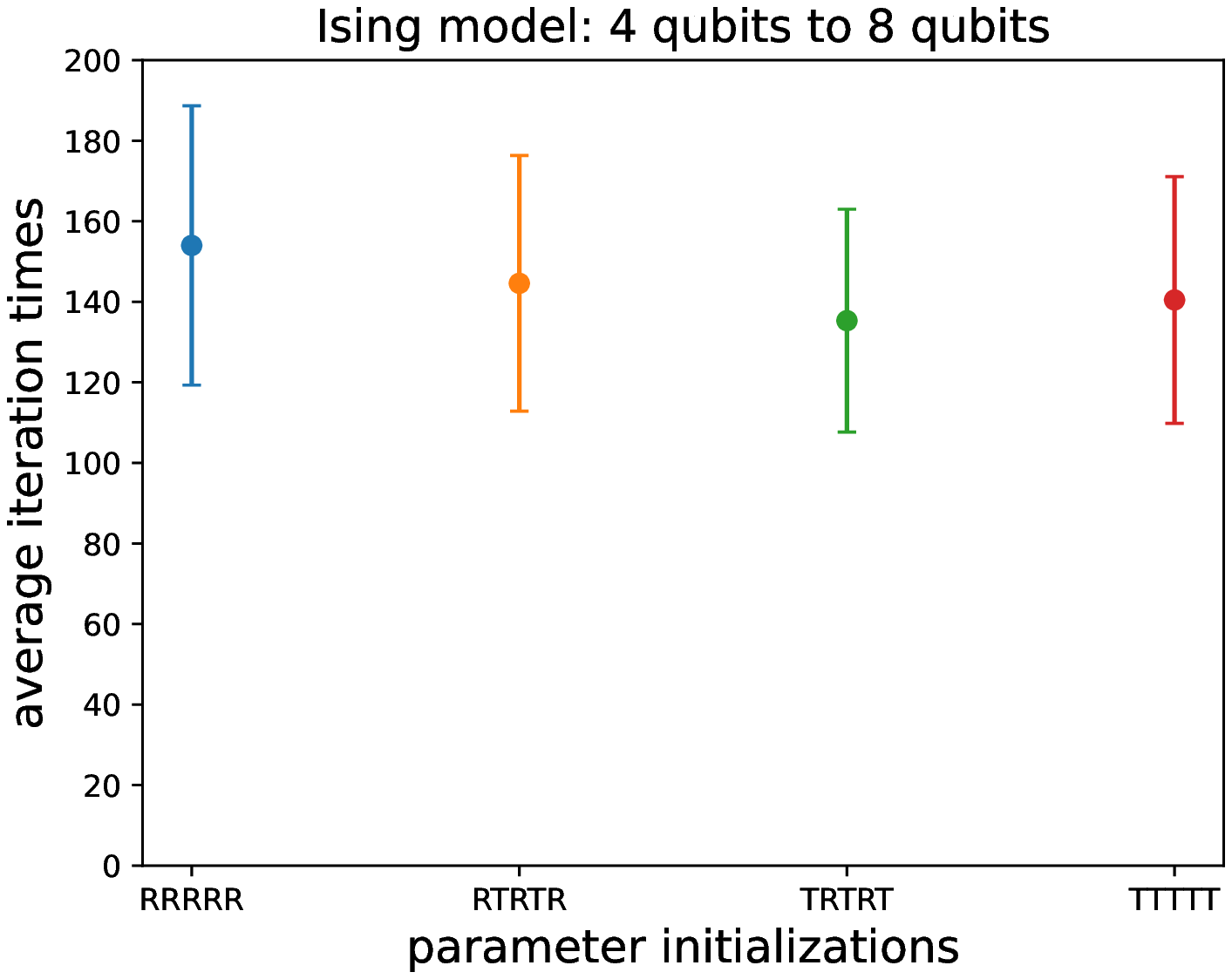}
  }
  \subfigure[\textbf{Task C}]{
  \includegraphics[width=0.3\linewidth]{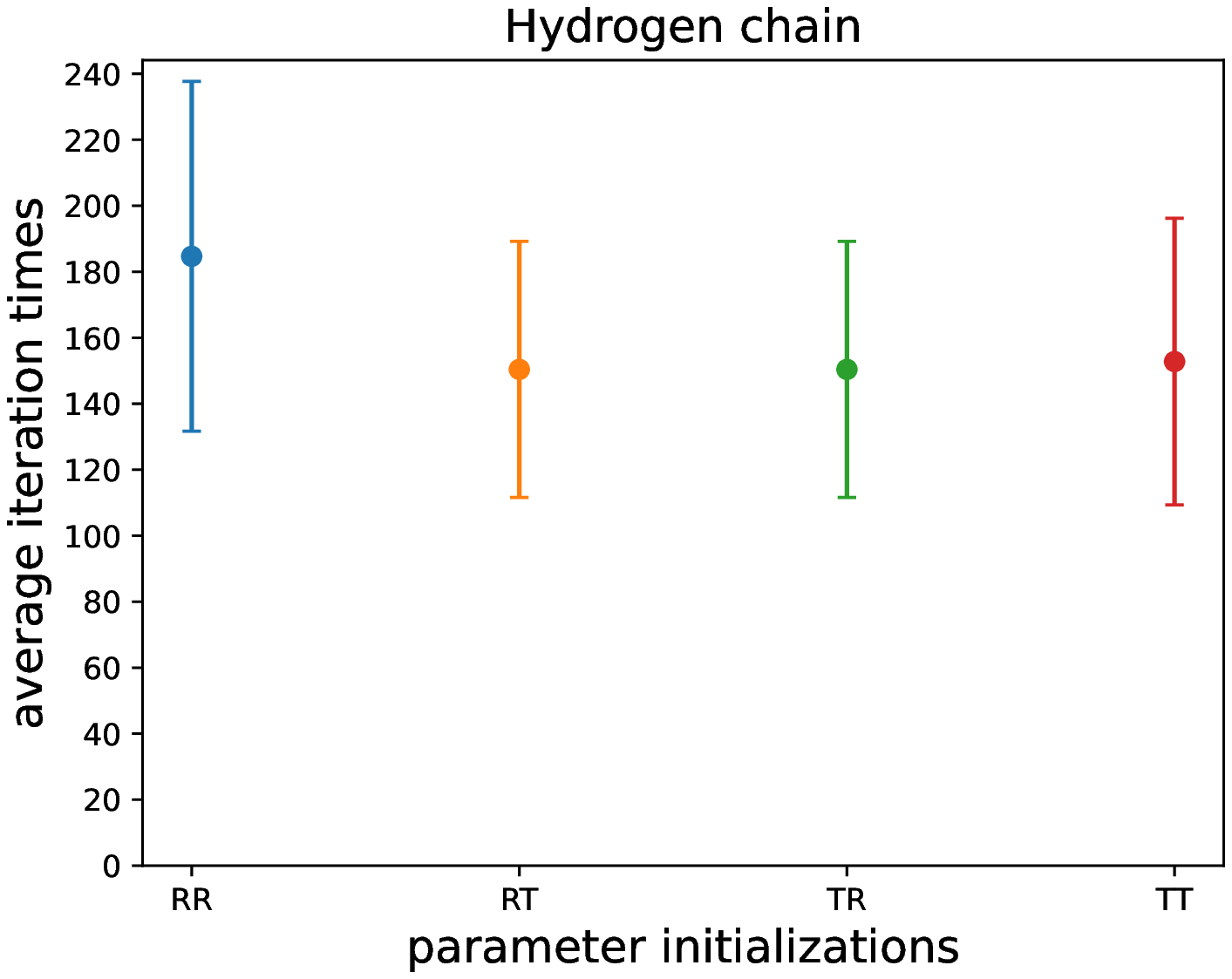}
  }
  \subfigure[\textbf{Task D}]{
  \includegraphics[width=0.3\linewidth]{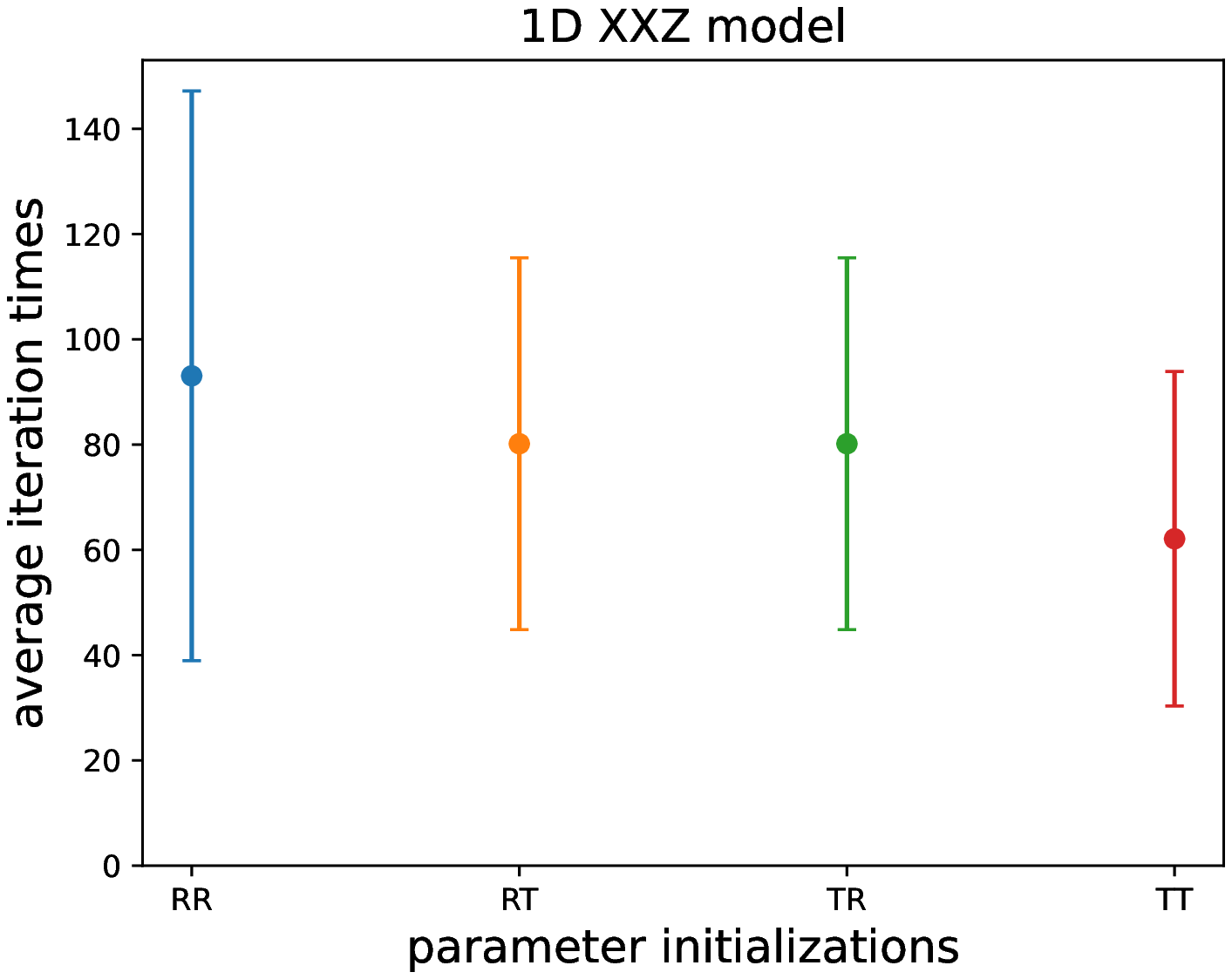}
  }
  \subfigure[\textbf{Task E}]{
  \includegraphics[width=0.3\linewidth]{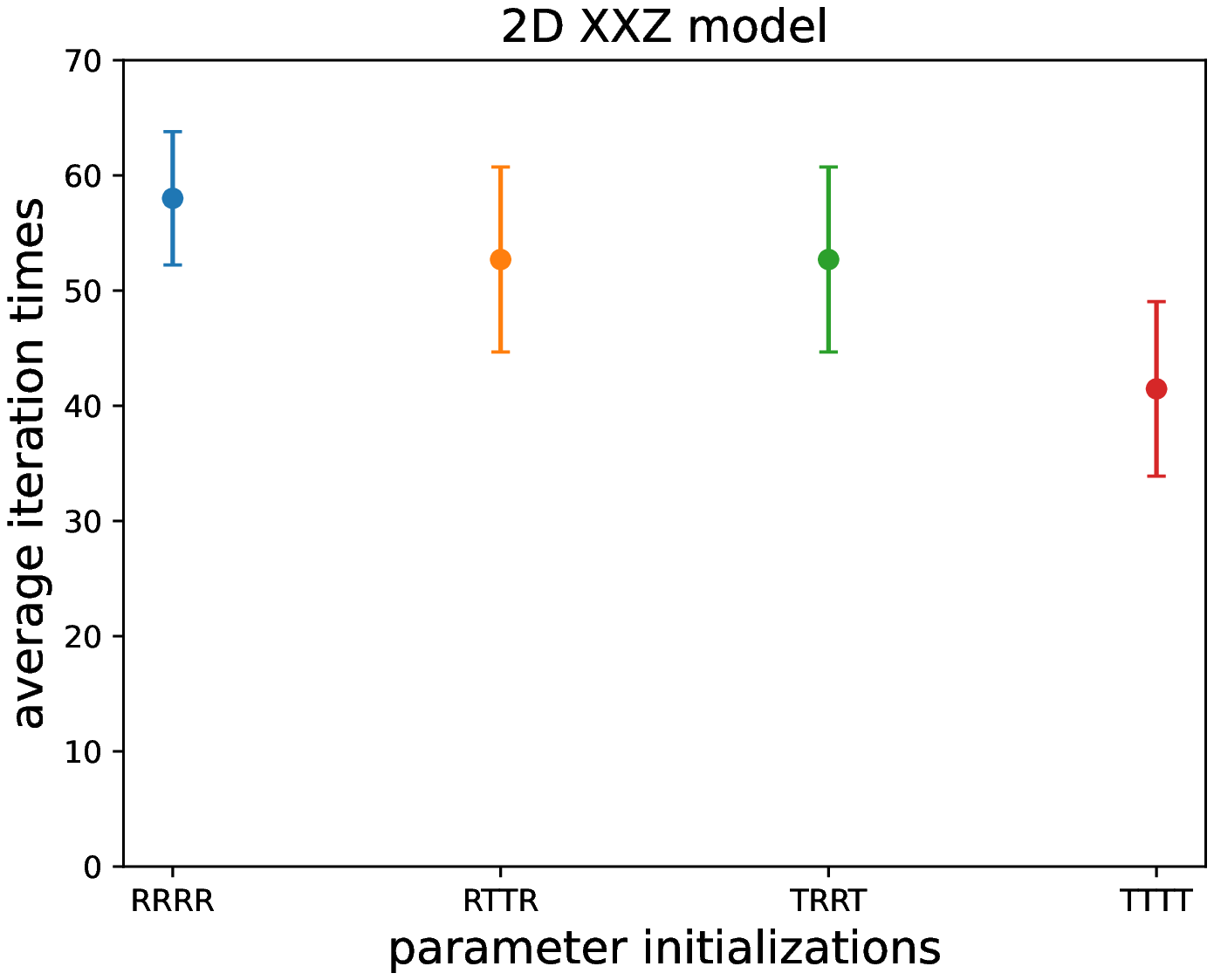}
  }
  \subfigure[\textbf{Task F}]{
  \includegraphics[width=0.3\linewidth]{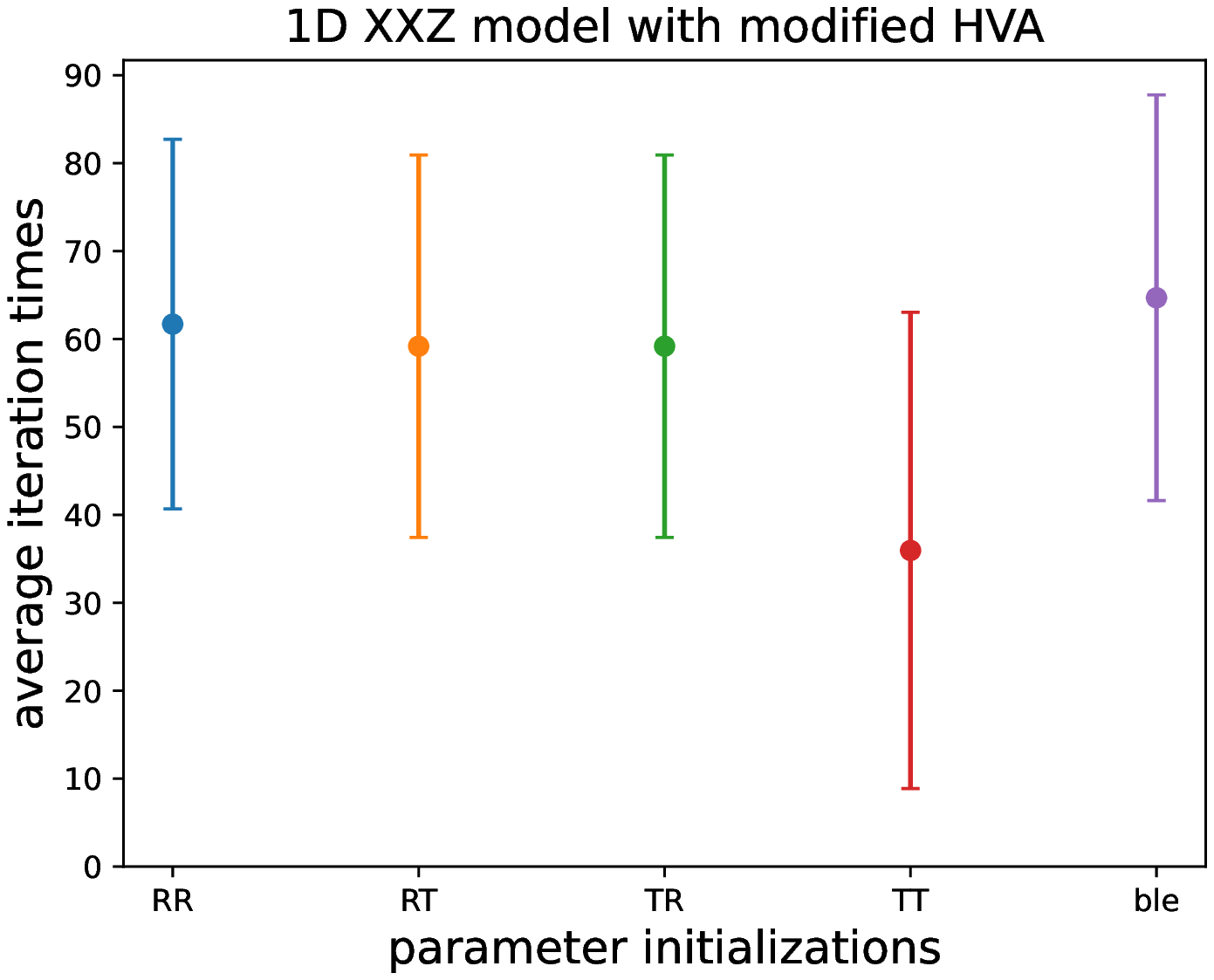}
  }
  \caption{
  \textbf{Average iteration times over 100 successful runs for different parameter initialization methods in every task. }Hot start parameters reduced this indicator, indicating that training efficiency is improved. All error bars represent standard deviations.
  }\label{ain}
\end{figure}

We plot the average iteration times for every task in Fig. \ref{ain}. In all tasks, hot-start parameters, especially the all ``T" initializations, reduced the average iteration times, showing that the method improved the training efficiency. Moreover, in most tasks, the standard deviations are also suppressed, meaning that simulations with transfer-learning-inspired initial parameters are more stable. Even though the standard deviations of all ``T" initializations in \textbf{Task E/F} are larger, the average value has been greatly reduced. Therefore, hot start parameters provide better initializations.

Finally, we consider the averaged normalized gradient norms, where the normalized gradient norm of a gradient $\{\partial_lC\}_{l=1}^L$ is: $G=\frac 1L   \sum_{l=1}^{L} (\partial_lC)^2 $. The average value determines the scaling of sampled gradients. Moreover, as shown in Lemma \ref{bpl}, when $\langle \partial_lC\rangle=0,\forall l$, the average value of $G$ is the average value of variances $\{\operatorname{Var}[\partial_lC]\}_{l=1}^L$.

\begin{figure}[ht]
  \centering
  \subfigure[\textbf{Task A}]{
  \includegraphics[width=0.4\linewidth]{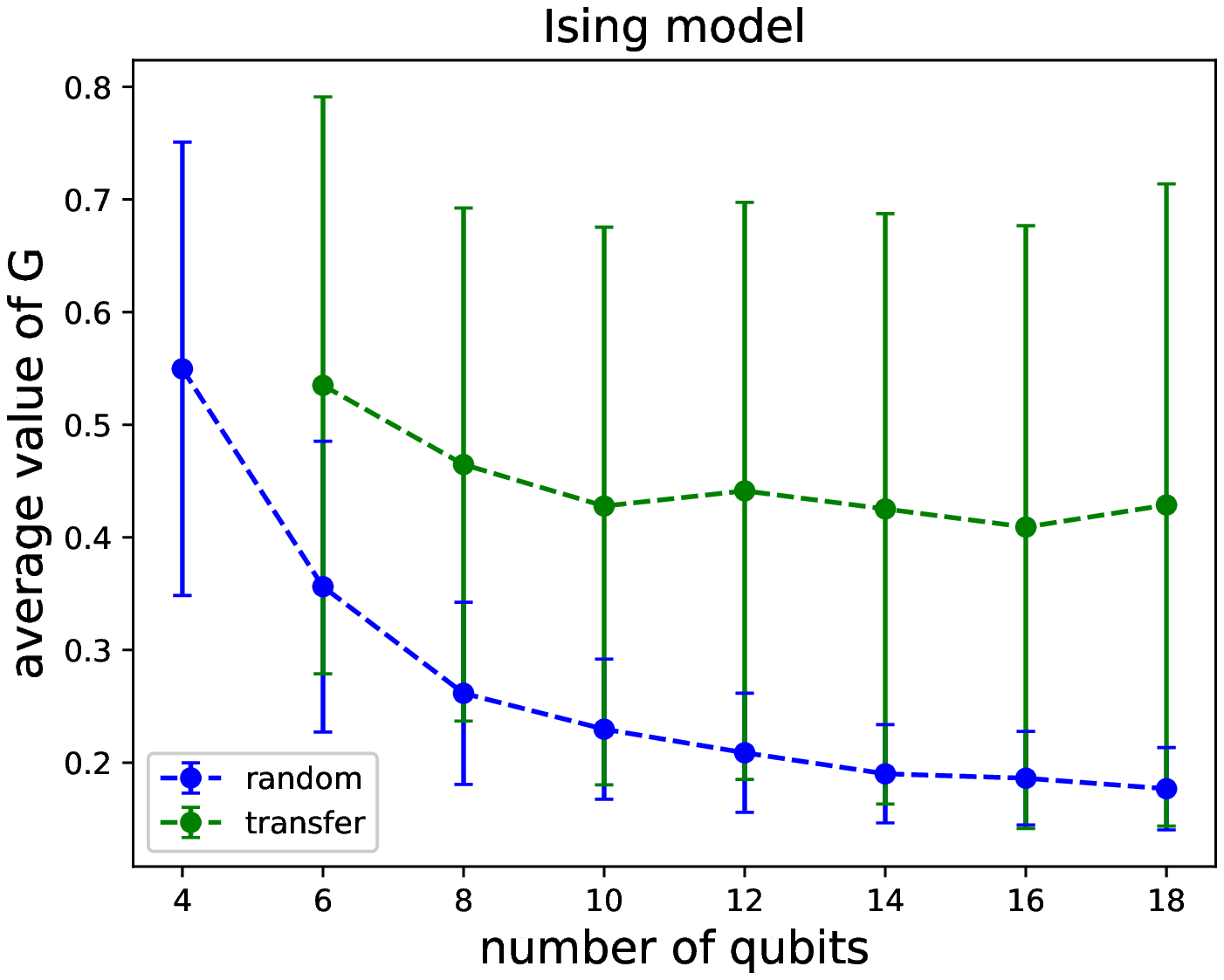}
  }
  \subfigure[\textbf{Task C}]{
  \includegraphics[width=0.4\linewidth]{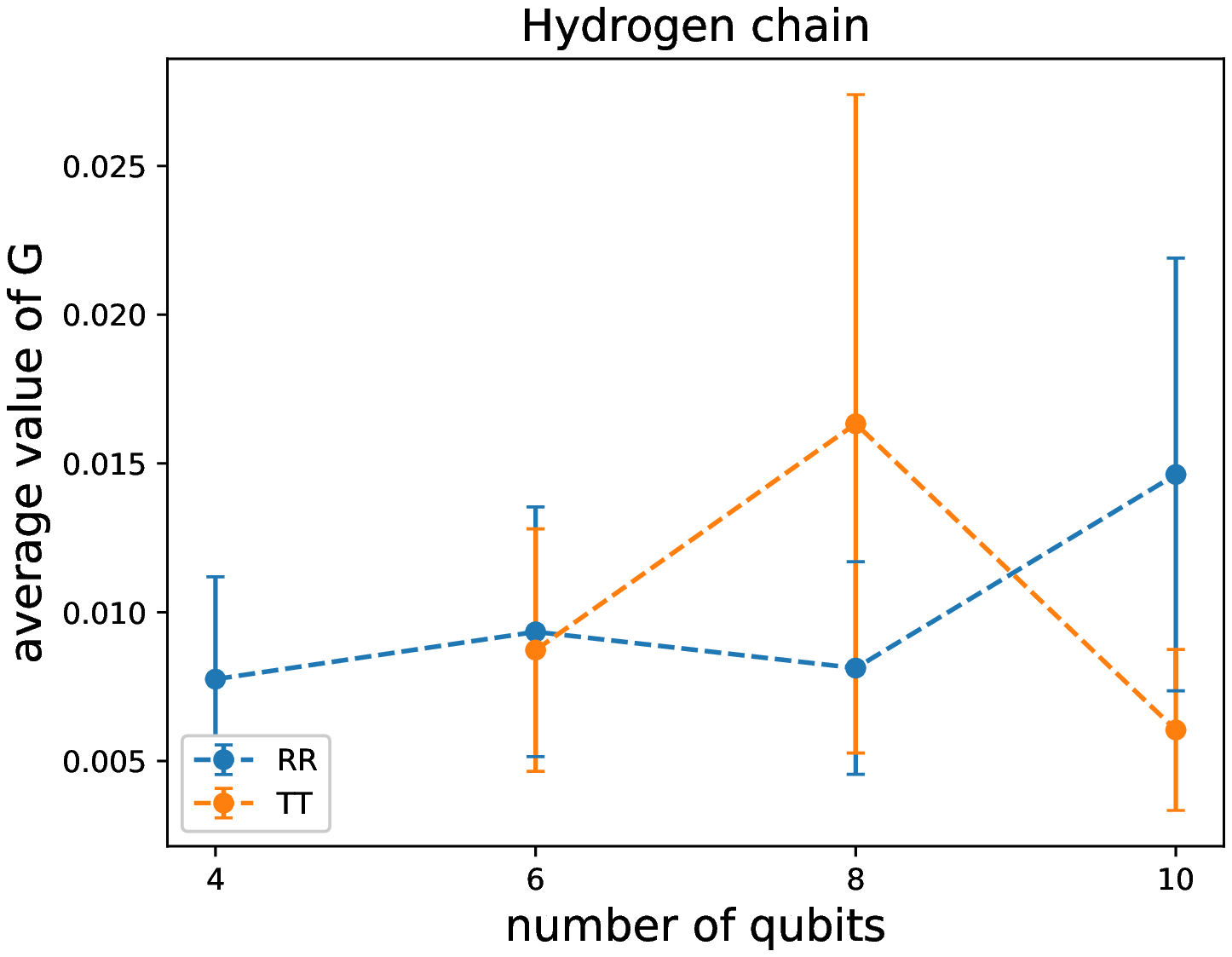}
  }
  \subfigure[\textbf{Task D}]{
  \includegraphics[width=0.4\linewidth]{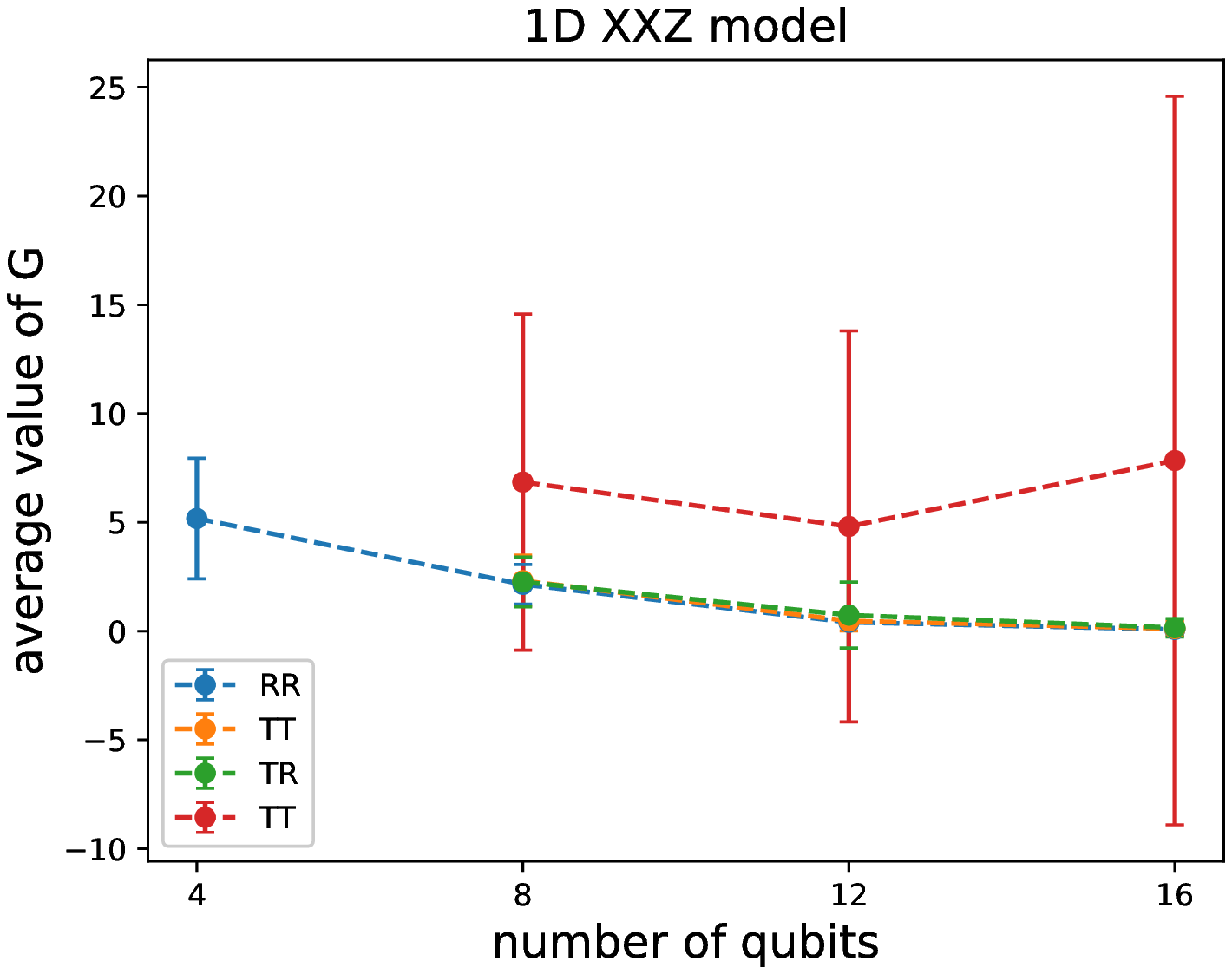}
  }
  \subfigure[\textbf{Task F}]{
  \includegraphics[width=0.4\linewidth]{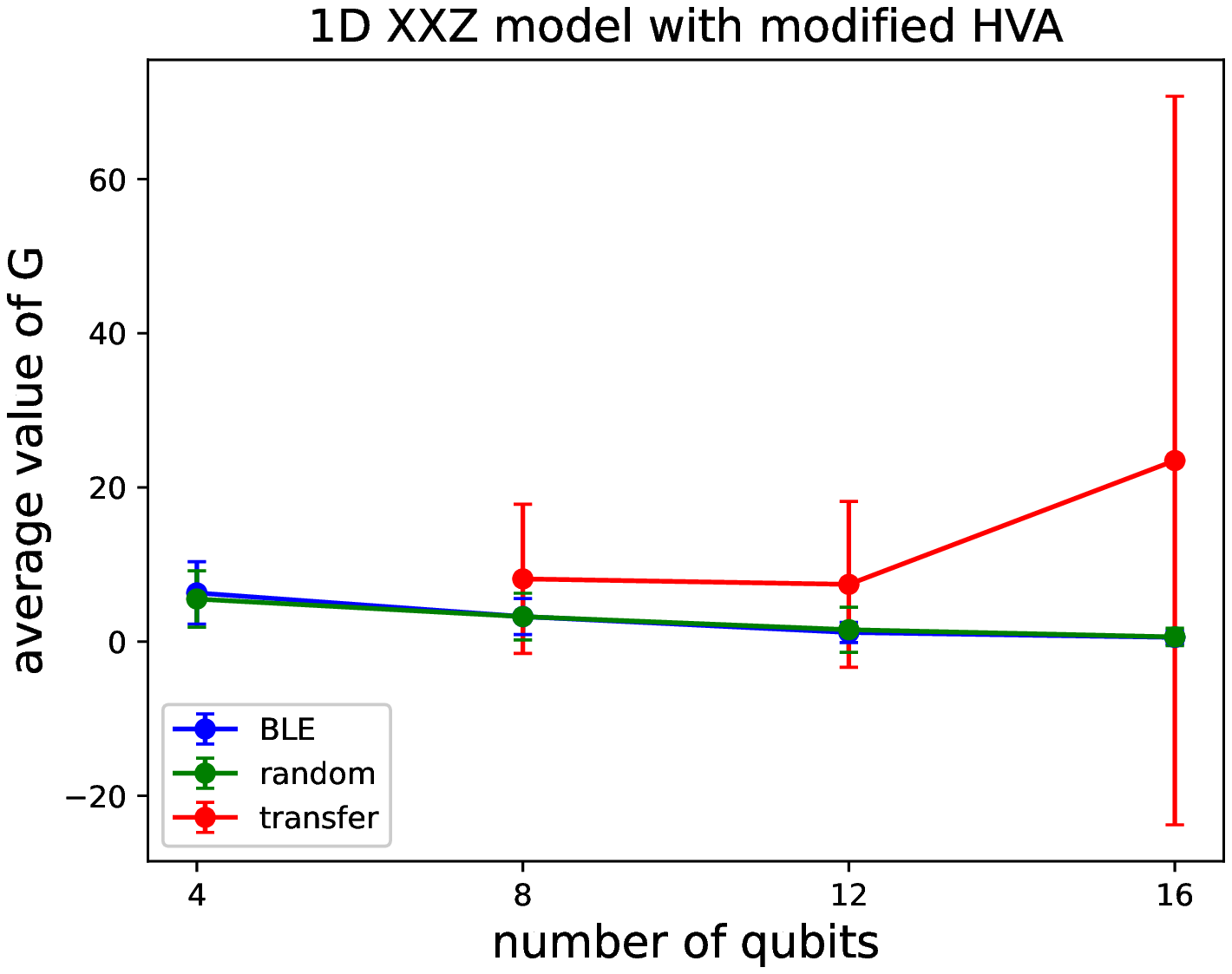}
  }
  \caption{
  \textbf{Average value of normalized gradient norms for different parameter initialization methods in some tasks. }Hot start parameters increased the average value and the standard deviations, indicating that BP is successfully mitigated.
  }\label{agn}
\end{figure}

Fig. \ref{agn} plots the average value of $G$ for different parameter initialization methods in some tasks. Hot start parameters can increase the scaling of gradients. In \textbf{Task C}, randomly initialized parameters do not decay all the time. The BLE strategy does not work well in \textbf{Task D}. Combining with former results, transfer-learning-inspired parameters mitigated BP and improved training efficiency.

Finally, an important phenomenon is that in most tasks, besides the cases where all parameters are randomly initialized or transferred from base tasks, labeled by ``RRR$\cdots$" and ``TTT$\cdots$", we also tested ``mixed strategies", like ``RTRT", where part of the parameters are transferred while the others are randomly initialized. The performance of such mixed strategies is better than the all ``T" initializations in some cases, especially when comparing the total trial number described in TABLE \ref{TTN}. The internal cause of this phenomenon may be that the blocks with randomly initialized parameters could help to escape from the local minima described by the solution of base tasks.

\section{Discussion}\label{discuss}

\subsection{Costs for training base circuits}
Note that whether training efficiency is improved is based on comparing the average iteration times and total trial number of different parameter initializations. However, when simulating hot start parameters, costs of training base circuits should be considered. Here, we will show that this overhead can be neglected, which does not affect simulation results. We consider the cost of evaluating one gradient vector via the parameter-shift rule \cite{psr} or the finite difference method:
\begin{equation}\label{tperstep}
  t_{\text{ per-step}} =2 \times  t_n \times L_n,
\end{equation}
where $t_n$ and $L_n$ are the time to evaluate the cost function and number of parameters when the qubit number is n.
On classical computers, simulation of quantum circuits costs exponentially with the number of qubits, then $  t_{\text{ per-step}}\in O(2^n)$.
On quantum computers, we assume the time to run a circuit is proportional to the number of layers $O(n)$. Then we have $  t_{\text{ per-step}}\in O( n^2 )$ for HVA and $  t_{\text{ per-step}}\in O( n^3 )$ for HEA.
It is also common that the number of iterations needed increases with the size of the system.
Combining these points with following simulation results, it is reasonable that the cost of training the base circuit can be neglected when comparing different parameter initialization methods.

\subsection{How can this method work well in general?}
While numerical simulation performed above showed that transfer-learning-inspired initial parameters can mitigate BP and improve training efficiency compared to randomly initialized ones, a natural question would be whether this method can work in general. We provide a discussion in this part.

First note that we demand the base task and target task, which differ in size, to be related. Therefore, applications in above numerical simulations are mainly focused on translation invariant systems, which is also the focus of this analysis. Other applications that refer to a generalized concept of ``related" can be inspired by classical machine learning tasks, which need to be further studied.

In this case, considering we are solving the ground state of the target system $H_{\text{target}}$, which can be divided into $K$ subsystems. To simplify our analysis, we assume $K=2$ and those subsystems are identical, each of which is an $n$-qubit Hamiltonian: $H_{\text{base}}^k,k=1,2$. The subscript ``base" indicates one subsystem is selected to be trained as the base task. Suppose we have successfully obtained its ground state: $|\psi_0\rangle^k_{\text{base}} =U^k_{\text{base}} (\bm{\theta}^*)|0\rangle^{\otimes n}$, with $\bm{\theta}^*$ optimum parameters. Then we group those subsystems:
$H_{\text{group}} = H_{\text{base}}^1 +  H_{\text{base}}^2 $.
Its ground state is:
\begin{equation}
|\psi_0\rangle_{\text{group}} = |\psi_0\rangle_{\text{base}}^1 \otimes |\psi_0\rangle_{\text{base}}^2 =  \left[  U^1_{\text{base}} (\bm{\theta}^*) \otimes U^2_{\text{base}} (\bm{\theta}^*)   \right] |0\rangle^{\otimes 2n }= U_{\text{group}} (\bm{\theta}^*) |0\rangle^{\otimes 2n}.
\end{equation}

Note that $H_{\text{target}}$ can be obtained as
\begin{equation}
H_{\text{target}} = H_{\text{group}} + H_{\text{interaction}},
\end{equation}
with $H_{\text{interaction}}$ the interaction term. Denote the ground state of $H_{\text{target}}$ as $|\psi_0\rangle_{\text{target}}$ and the fidelity between $|\psi_0\rangle_{\text{target}}$ and $|\psi_0\rangle_{\text{group}}$ is $F_1$. Generally, $F_1$ decreases with the interaction strength between subsystems.

On the other hand, when training the target task, the target circuit is also slightly modified based on $U_{\text{group}} (\bm{\theta}^*) $.  For instance, in HEA, entangling operations in one layer are changed as:
\begin{equation*}
\left[  \prod_{i=1}^{n-1} CZ_{i,i+1}  \right] CZ_{n,1} \otimes \left[  \prod_{i=n+1}^{2n-1} CZ_{i,i+1}  \right] CZ_{2n,n+1} \to \left[  \prod_{i=1}^{2n-1} CZ_{i,i+1}  \right] CZ_{2n,1}.
\end{equation*}

Denote the modified target circuit with transferred initial parameters as $U_{\text{target}}(\bm{\theta}^*)$. Denote the fidelity between $U_{\text{target}}(\bm{\theta}^*) |0\rangle^{\otimes 2n}$ and $|\psi_0\rangle_{\text{group}}$ is $F_2$. Since we only made small changes in the circuit, then there exist parameters that can lead to high-valued $F_2$.

Finally, we focus on the fidelity between $|\psi_0\rangle_{\text{target}}$ and $U_{\text{target}}(\bm{\theta}^*) |0\rangle^{\otimes 2n}$. Note that a larger fidelity indicates that the initial state of the target circuit is closer to the ground state, which requires fewer costs for training. Moreover, a larger fidelity indicates a lower cost function. which could be located out of the BP region with a high probability according to the cost concentration introduced before.

Therefore, a larger fidelity is preferred. Obviously, the fidelity increases with both $F_1$ and $F_2$. Increasing $F_1$ and $F_2$ could improve the performance of this method. This conclusion guides how our method could work well. On the one hand, larger $F_1$ means a weaker interaction strength between subsystems. On the other hand,  before training the target circuit, we can compute $F_2$, which can be efficiently evaluated, and train the target circuit with the set of parameters corresponding to a high-valued $F_2$.

The results above are based on $K=2$ and those subsystems are identical. When $K>2$, a similar analysis can be performed. When those subsystems are not identical, we can select and train multiple subsystems as base tasks.

\section{Conclusion}\label{conclusion}
In this paper, inspired by transfer learning, we proposed a parameter initialization method for VQAs. Information from a pre-trained small-sized task is further applied in larger-sized tasks. The structure transfer method and network transfer method are introduced to transfer the information. Numerical simulations showed that this method can mitigate BP and improve training efficiency compared to randomly initialized parameters.

We also provided a discussion on how this method can work in general. Results showed that when the following conditions are satisfied, our method could work well: a) The target system can be divided into several individual parts without too many changes in the eigen-information. b) The overlap between the grouped optimum base ansatz and the transferred initial one is high. Therefore, current applications of the method are mainly focused on translation invariant systems. Other applications need to be further studied, which may be inspired by classical machine learning.

Two main points should be addressed here. The first is that the main focus of this paper is to improve the trainability of VQAs with the proposed parameter initialization method. However, whether optimizing the ansatz will succeed in finding the solution is not straightforward. Such a topic is in the context of reachability, where little results have been provided \cite{qaoalayer,reach,uvqa}. Further study is required on this topic.

The second point is that the problem of interpretability and transferability in transfer learning is still wide open. Our discussion is based on translation invariant systems. General performances of this method need to be studied. However, as shown in \cite{expbp}, we applied correlated parameters instead of randomly initialized ones. Then improvements in the trainability are guaranteed.

\acknowledgments
We thank Zhao-Yun Chen and Cheng Xue for helpful discussions.
This work was supported by the National Natural Science Foundation of China (Grant No. 12034018), and Innovation Program for Quantum Science and Technology No. 2021ZD0302300.
The numerical calculations in this paper have been done on the supercomputing system in the Supercomputing Center of University of Science and Technology of China.

\section*{Data availability}
The data that support the findings of this study are available upon reasonable request from the authors.

\bibliography{ref}

\end{document}